\documentclass[12pt]{article}

\title{\bf
                Remote control system of\\
                a binary tree of switches --\\
                II. balancing for a perfect binary tree
}

\author{
  O.~Golinelli\\
  \normalsize\itshape Institut de Physique Th\'{e}orique,
  \normalsize\itshape CEA, CNRS, Universit\'{e} Paris-Saclay, France
}
\date{}

\usepackage[a4paper, left=30mm, right=30mm, top=30mm, bottom=34mm]{geometry}

\usepackage[final]{graphicx}
\graphicspath{{../figures/pdf}}

\usepackage[final, breaklinks, colorlinks, allcolors=black]{hyperref}
\begin{document}

\maketitle

\begin{abstract}
%---------------
  \normalsize
We study a tree coloring model introduced by Guidon (2018), initially based
on an analogy with a remote control system of a rail yard, seen as switches
on a binary tree.  For a given binary tree, we formalize the constraints on
the coloring, in particular the distribution of the nodes among colors.
Following Guidon, we are interested in balanced colorings i.e.\ colorings
which minimize the maximum size of the subsets of the tree nodes
distributed by color.  With his method, we present balanced colorings for
trees of height up to 7.  But his method seems difficult to apply for trees
of greater height.
Also we present another method which gives solutions for arbitrarily large
trees.  We illustrate it with a balanced coloring for height 8.
In the appendix, we give the exact formulas and the asymptotic behavior of
the number of colorings as a function of the height of the tree.
\end{abstract}

%===========================================================================

\section{Introduction}
%---------------------

Since the appearance of computers, the notion of tree is central in
computer sciences.  For example to decompose program lines, the first
Fortran compilers used a binary tree optimized for keyword recognition
time.  Tree structures are intensively used for databases, algorithms,
representation of expressions in symbolic programming languages, etc.

Many variants have been developed to minimize searching time or to save
memory, like B-tree, red–black tree, etc.  Trees are essential for network
design and parallel computing: many computer clusters have a fat-tree
network and algorithm performances depend on communication links
between computer nodes.

The mathematician and computer scientist D.~ Knuth, also creator of the TeX
typesetting system, devotes a hundred pages to trees in his encyclopedia,
\emph{The art of computer programming}, and summarizes their development
since 1847~\cite[page 406]{Knuth}.

By leafing through a general public review on Linux, we read an
article~\cite{Guidon-1} that describes how to control a binary tree of
electronic switches with a minimum number of signals.  The
author Y.~Guidon explains how to balance the signal power to avoid the bad
situation where a single signal controls half of the switches.  For that, he
describes this problem in terms of binary tree coloring, but with a rule
different from that usual in graph theory.  He draws balanced
solutions~\cite{Guidon-1, Guidon-2} for perfect binary trees with height up
to 5 with a heuristic algorithm.

Our article is devoted to a systematic algorithm for coloring a tree of
arbitrary height.  In Section~\ref{sec:definitions}, we recall some usual
definitions on trees.  As Ref.~\cite{Guidon-1} is in French and difficult
to access for non-subscribers, we recall its coloring rule in
Section~\ref{sec:color}, the notion of colorable partitions in
Section~\ref{sec:partitions}, and of balanced coloring in
Section~\ref{sec:balanced}.
In Section~\ref{sec:algorithm}, we describe a systematic algorithm to color
a perfect binary tree, whatever its height, with a given colorable
partition, balanced or not.  Then we end by giving examples in
Section~\ref{sec:examples}.  Some details are reported in Appendix.

%===========================================================================

\section{Tree coloring}
%----------------------

\subsection{Definitions}
\label{sec:definitions}

In this article, we only consider \emph{perfect binary trees} (sometimes
also called \emph{complete binary trees} or \emph{full binary trees}) of
height $h$ ($h \ge 0$ and integer).
We give a recursive definition:
a perfect binary tree $T$ of height $h$ decomposes into $T = (r, T_L, T_R)$
where $r$ is a \emph{node}, called \emph{root}, $T_L$ and $T_R$, left and
right subtrees, are perfect binary trees of height $h-1$.
In the particular case $h=0$, $T$ is reduced to its root $r$.

We can represent a tree by a connected graph without cycle, with a marked
node, the root $r$.  Let $u$ and $v$ be different nodes; we say that $u$
is an \emph{ancestor} of $v$, or equivalently $v$ is a \emph{descendant} of
$u$, if $u$ is on the shortest path between $r$ and $v$.
The root is the only node without an ancestor.  In contrast, the
\emph{leaves} are defined as the nodes without descendants.

The \emph{height} of a node $u$ is defined as the distance between $u$
and $r$.  In a perfect binary tree,
the leaves are all the same height, which is the height of the tree.
In a perfect binary tree of  height $h$, there are $2^i$ nodes of height
$i$, with $0 \le i \le h$; then a total of $2^{h+1}-1$ nodes.

As in this article we only consider perfect binary trees, we use the word
\emph{tree} for perfect binary tree by simplification.

\subsection{Coloring rule}
\label{sec:color}

As explained in Refs.~\cite{Guidon-1,Guidon-2,og-1}, the initial problem of
switches can be described with graph coloring terminology.
For a given graph, a coloring consists in assigning a label (called
\emph{color}) to each node of the graph, respecting a rule.

In graph theory, the usual rule is that two directly connected nodes have
different colors.  But this rule is of no interest for trees.  As a tree is
a graph without loop, it can be colored with only two colors by alternating
them along each path.  Also we impose a stronger rule:

\medskip\noindent\textbf{Coloring rule:}
\label{sec:rule}
\emph{For a tree $T$ and each pair of nodes $\{u,v\}$ of $T$,
if $u$ is an ancestor of $v$, then $u$ and $v$ have different colors.}

It is equivalent to the following rule:
\emph{for each leaf $v$ of $T$, nodes along the path from the root of $T$
  to $v$ are all different colors.}

Consequently, to color a tree of height $h$, we need at least $h+1$ colors,
because the path from root to any leaf includes $h+1$ nodes.

\begin{figure}
   \centering
   \includegraphics{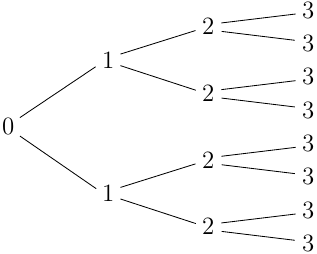}
   \caption{Canonical coloring of a tree with height 3.}
   \label{fig:h3}
\end{figure}

The simplest coloring is the \emph{canonical coloring by height}: the
nodes of height $i$ have the color $i$, for $0 \le i \le h$, therefore
$h+1$ colors.  See Fig.~\ref{fig:h3}.

Thus, for a tree of height $h$, it is always possible to color it with
$h+1$ colors and this is the minimum number of colors to respect the
coloring rule.
This does not prohibit coloring it with more colors.  For example, we can
use a different color for each node, so $2^{h+1}-1$ colors.
In this article, we only consider colorings with the minimum number of
colors, i.e.\ $h+1$ colors.

\begin{figure}
   \centering
   \includegraphics{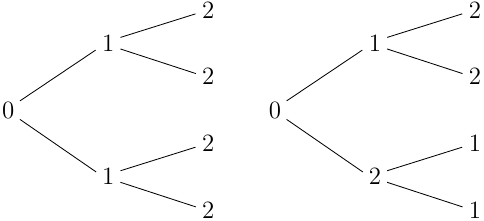}
   \caption{The two non-equivalent colorings of a tree of height 2 with 3
     colors, the canonical one (left) and another (right).}
   \label{fig:h2}
\end{figure}

Of course colorings other than the canonical coloring are possible.
As values of color label do not matter, we can
change or permute the labels without effects; only counts the partition of
the tree into subsets of nodes having the same color.  Then two colorings are
equivalent if they give the same subsets of nodes.
For example, for a tree of height 2, there are only 2 non-equivalent colorings
with 3 colors~\cite{Guidon-1}; see Fig.~\ref{fig:h2}.

By enumeration, we can see that there are 24 different colorings for a tree
of height 3 with 4 colors, and $24^3 = 13824$ for height 4 with 5 colors.

Let $c_n$ be the number of different colorings for a tree of height $n-1$
with $n$ colors; we give Eq.~(\ref{cn}) for $c_n$ in Appendix.
Its asymptotic expansion for large $n$ begins with
\begin{equation}
    c_n = \frac{U^{2^n}}{n \: n!} (1 - 2/n + 5/n^2 - 16/n^3 + 66/n^4
                                  - 348/n^5 + \ldots) ,
\end{equation}
where $U = 1.6616879496\ldots$ is the constant given in Eq.~(\ref{u}).

We can interpret $\log c_n$ as a coloring entropy.
As a tree of height $n-1$ has $2^n-1$ nodes, the entropy per
node $(\log c_n) / (2^n-1)$ is asymptotically
\begin{equation}
    \sigma = \log U = \sum_{k \ge 2} \frac{\log k}{2^k}
           = 0.5078339228\ldots  \label{sigma}
\end{equation}
whose digits form the sequence A114124 in OEIS~\cite{OEIS}.
In other words, when we color a large tree, the number of colorings
increases as if there are on average $U$ choices to color each node.

As we have $n$ choices to color the root, then $n-1$ choices to color the
children of the root, etc. one may be surprised that the entropy per node
is asymptotically a constant and does not increase with $n$.
But notice that at the last step of the coloring process, the color of each
leaf is imposed; the leaves represent asymptotically $1/2$ of the nodes of
the tree and have zero entropy.
Similarly for the parents of the leaves which
represent asymptotically $1/4$ of the nodes of the tree,
there are only 2 choices per node.
More generally, there are $k$ choices to color each node of height $n-k$;
these nodes form asymptotically $1/2^k$ of the nodes. The sum over $k$ gives
Eq.~(\ref{sigma}) for $\sigma$.

%===========================================================================

\section{Colorable partitions}
\label{sec:partitions}

Let $T$ be a perfect binary tree of height $h$ with a coloring that
respects the rule defined in Section~\ref{sec:rule} and with the minimum
number $h+1$ of colors.  The colors are indexed by $i = 0, \ldots, h$ and
we denote by $a_i$ the number of nodes of $T$ of color $i$.

We have the sum rule $\sum_i a_i = N_h$, where $N_h = 2^{h+1}-1$ is the
number of nodes of $T$.  Moreover, as $h+1$ is the minimum number of colors,
there is at least one node of each color, hence $a_i \ge 1$, for all $i$.

In other words, $A = (a_0, a_1, \ldots, a_h)$ is a partition of the integer
$N_h$; we say $A$ is a \emph{colorable partition} of $T$.
As two coloring are equivalent by color permutation,
the order of the $a_i$ does not matter.
By convention, the partitions are written in non-decreasing order.

For example, the colorable partition for the tree in Fig.~\ref{fig:h3} is
$A=(1,2,4,8)$.
More generally, for a tree of height $h$, its \emph{canonical} coloring
gives the colorable partition $A_c = (1,2,4 \ldots, 2^h)$ with $a_i=2^i$.

For a tree of height $h=2$, there are two possible colorings drawn in
Fig.~\ref{fig:h2} which give two different colorable partitions, the canonical
one $(1,2,4)$ and another $(1,3,3)$.

For a tree of height $h=3$, there are 24 possible colorings, which give
only 8 colorable partitions, which are $(1,2,4,8)$,
$(1,2,5,7)$, $(1,2,6,6)$, $(1,3,3,8)$, $(1,3,4,7)$, $(1,3,5,6)$,
$(1,4,4,6)$, $(1,4,5,5)$.
Consequently, a colorable partition can sometimes correspond to several
different colorings.

\begin{figure}
   \centering
   \includegraphics[width=\textwidth]{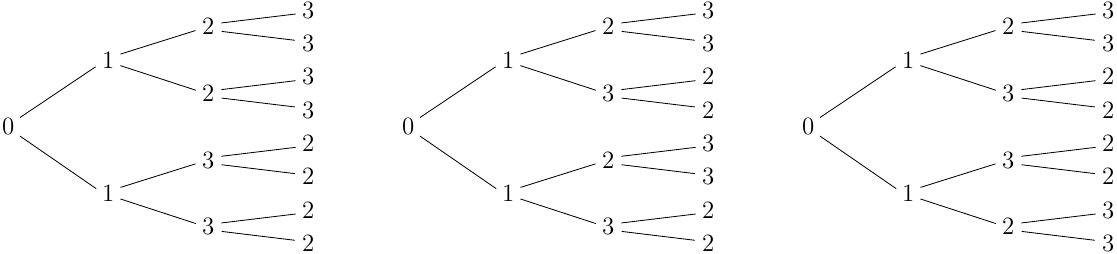}
   \caption{The 3 colorings of a tree of height 3 with colorable partition
     $A = (1,2,6,6)$.  For each coloring, there are one node with color 0,
     two nodes with color 1, six nodes with color 2 and six nodes with color 3.}
   \label{fig:a1266}
\end{figure}

For example, for $h=3$, the colorable partition $A = (1,2,6,6)$ corresponds
to 3 colorings, represented on Fig.~\ref{fig:a1266}.  The two colorings
on the right are deduced from each other by a up-down symmetry in one of the
subtrees attached to the root; a spatial symmetry in any subtree does not
change the distribution of colors.  On the other hand, the coloring on the
left cannot be deduced from the others by a simple symmetry.

Conversely, the partitions of the integer $N_h$ into $h+1$ positive
integers do not always correspond to a coloring of a tree of height $h$.
For example, for $h=3$, there are 27 partitions of the integer $N_3=15$
into 4 parts, but only the 8 partitions indicated above are colorable
partitions.

In Ref.~\cite{og-1}, we show for any rooted tree (binary or not), what are
the constraints on a partition so that it can be colorable.
In the particular case of the perfect binary trees, these constraints can be
written, for a tree of height $h$ and a colorable partition
$A = (a_o, a_1, \ldots, a_h)$:
\begin{equation}
  \sum a_i = 2^{h+1}-1, \label{sum}
\end{equation}
\begin{equation}
  a_i = 1 ; a_j \ge 2 \label{a1}
\end{equation}
for one and only one color $i$ (the color of the root), and  $j \ne i$,
and moreover for any subset of $k$ different colors
$\{i_1, i_2 \ldots, i_k\}$ with $1 \le k \le h+1$,
\begin{equation}
  a_{i_1} + a_{i_2} + \cdots + a_{i_k} \ge 2^k-1 .  \label{ak}
\end{equation}

The reader can find all the details of the demonstration of these relations
in Ref.~\cite{og-1}, valid for any rooted tree.
Notice that in Ref.~\cite{og-1},
the inequalities are written with an \emph{upper} bound
\begin{equation}
  a_{i_1} + a_{i_2} + \cdots + a_{i_k} \le 2^h + 2^{h-1} + \cdots + 2^{h-k+1}
                                     = 2^{h+1} -  2^{h+1-k} . \label{akinv}
\end{equation}
But there is no contradiction with Eq.~(\ref{ak}): by complementarity,
using the sum rule Eq.~(\ref{sum}), we find the \emph{lower} bound $2^{k'}-1$
for the $k' = h+1-k$ other colors.  Therefore, Eqs.~(\ref{ak})
and~(\ref{akinv}) are equivalent.

In Ref.~\cite{og-1}, we have shown that the above conditions are necessary,
but not always sufficient for any rooted tree.
But for perfect binary trees, thanks to their numerical balance between all
its branches, these necessary conditions are also sufficient.
This last point is demonstrated by the recursive algorithm, exposed
Section~\ref{sec:algorithm},
which constructs a coloring from any partition which obeys the
conditions Eqs.~(\ref{sum}--\ref{ak}).

%===========================================================================

\section{Balanced coloring}
%----------------------------
%
\label{sec:balanced}

We saw in the previous section that there are many possible distributions
of colors for a tree of given height $h$, and respecting the coloring rule.
In particular, canonical coloring by height gives a partition
$A_c = (1,2,4,\ldots, 2^h)$.
As the colorable partitions must respect the
conditions Eqs.~(\ref{sum}-\ref{ak}), it turns out that $A_c$ is the
maximally unbalanced partition;
more than half of the nodes have the same color, 3/4 use 2 colors,
etc. And more generally, it is the only coloring where the lower or upper
bounds of Eqs.~(\ref{ak}, \ref{akinv}) are reached for all $k$.

Y.~Guidon~\cite{Guidon-1} has studied how to best balance the coloring
of a tree.
Here the optimization work relates only to the coloring, keeping the tree
graph fixed.
As $\sum_i a_i = N_h$, the number of nodes, is fixed, the optimization
criterion relates to another quantity, for example to minimize
$\sum_i a_i^2$, or more generally the $p$'th moment $\mu_p = \sum_i a_i^p$, or
the entropy
$S = - \sum_i a_i \log(a_i) = - \left. d \mu_p/dp \right|_{p=1}$.
Following Guidon~\cite{Guidon-1}, we choose the simplest criterion:
minimize $\max_i(a_i)$, which corresponds to the minimization of $\mu_p$ for
large $p$.

For a tree of height $h$, therefore with $N_h=2^{h+1}-1$ nodes, colored
with $h+1$ colors, the solution is given by Guidon~\cite{Guidon-1}.  We
repeat it here.

First we have to choose a color, for example color 0 for the root only, so
$a_0=1$.  For the $h$ other colors, we try to distribute them in a balanced
way on the $m$ other nodes, with $m=N_h-1= 2^{h+1}-2$.
This gives $a_i = m/h$ nodes
per color when $m$ is divisible by $h$.  Otherwise, the remainder $r$ of
the integer division must be distributed among $r$ colors.

Let $q$ and $r$ the quotient and the remainder of the Euclidean division
$m/h$, such that  $m = hq+r$ with $0 \le r < h$.
The balanced partition is $A_b = (1, q, q, \ldots, q+1, q+1, \ldots)$ with
$h-r$ colors with $q$ nodes and $r$ colors with $q+1$ nodes.
We can verify that the partition $A_b$ satisfies the
conditions Eqs.~(\ref{sum}-\ref{ak}).
Balanced partitions for small heights are given in
Table~\ref{table:ab}.
For large $h$, the asymptotic behavior is $\max_i a_i \sim 2^{h+1}/h$.

% Table
%      0   1       1
%      1   3       1 2
%      2   7       1 3 3
%      3   15      1 4 5 5
%      4   31      1 7 7 8 8
%      5   63      1 12*3 13*2
%      6   127     1 21*6
%      7   255     1 36*5 37*2
%      8   511     1 63*2 64*6
%      9  1023     1 113*4 114*5
%     10  2047     1 204*4  205*6
%     11  4095     1 372*9  373*2

\begin{table}
   \centering
   \begin{tabular}{|rrl|}
      \hline
       h  &   $N_h$  & $A_b$ \\
       \hline
       0  &   1      &  (1) \\
       1  &   3      &  (1, 2) \\
       2  &   7      &  (1, 3, 3) \\
       3  &   15     &  (1, 4, 5, 5) \\
       4  &   31     &  (1, 7, 7, 8, 8) \\
       5  &   63     &  (1, $12^3$, $13^2$) \\
       6  &   127    &  (1, $21^6$) \\
       7  &   255    &  (1, $36^5$, $37^2$) \\
       8  &   511    &  (1, $63^2$, $64^6$) \\
       9  &  1023    &  (1, $113^4$, $114^5$) \\
      10  &  2047    &  (1, $204^4$,  $205^6$) \\
      11  &  4095    &  (1, $372^9$,  $373^2$) \\
      \hline
   \end{tabular}
   \caption{Balanced colorable partitions for perfect binary trees of
     height $h$ with $N_h=2^{h+1}-1$ nodes, colored with $h+1$ colors,
     published by Guidon~\cite{Guidon-1}.  The notation $x^k$ is for $k$
     repetitions of $x$.
   }
   \label{table:ab}
\end{table}

We described above how to find the values $a_i$ for the balanced colorable
partition $A_b$, but
the real difficulty is to find how to color a tree with precisely this
partition $A_b$.

For small values of $h$, a solution is easily found by trial and error.
For example, with $h=2$, the coloring on the left on Fig.~\ref{fig:h2} is
well balanced.  For intermediate values, we will still find a solution, but
with more difficulty. But if we want solutions for large $h$, we need a
systematic algorithm.

\begin{figure}
   \centering
   \includegraphics{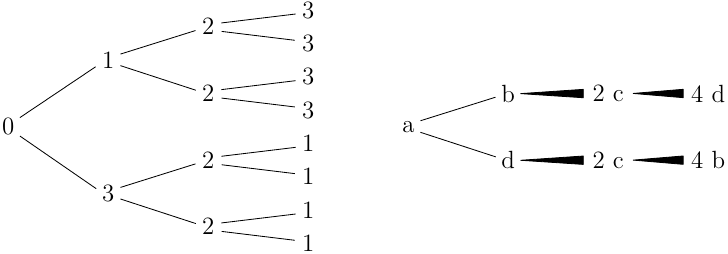}
   \caption{A balanced coloring $(1,4,5,5)$ given by Guidon~\cite{Guidon-1}
     for a tree with height 3.  The drawing on the right is a compact
     representation of the one on the left. For convenience, the colors are
     now labeled by letters $a,b,c,d$ and a notation like $b-2c-4d$
     represents a subtree with a canonical coloring where each $2^k$ is
     the number of nodes with the same height and the same color: one node
     colored with $b$, 2 nodes with color $c$ and 4 nodes with color $d$.
   }
   \label{fig:g3}
\end{figure}

\begin{figure}
   \centering
   \includegraphics[width=\textwidth]{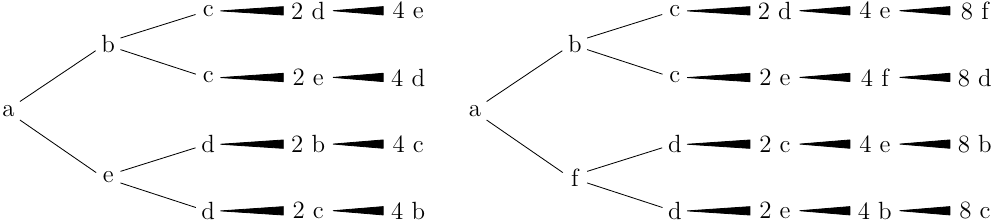}
   \caption{Left: a balanced coloring $(1,7,7,8,8)$ for a tree with height 4.
     Right: a balanced coloring $(1,12,12,12,13,13)$ for a tree with height 5.
     These two solutions are given by Guidon~\cite{Guidon-1}.
   }
   \label{fig:g45}
\end{figure}

\begin{figure}
   \centering
   \includegraphics{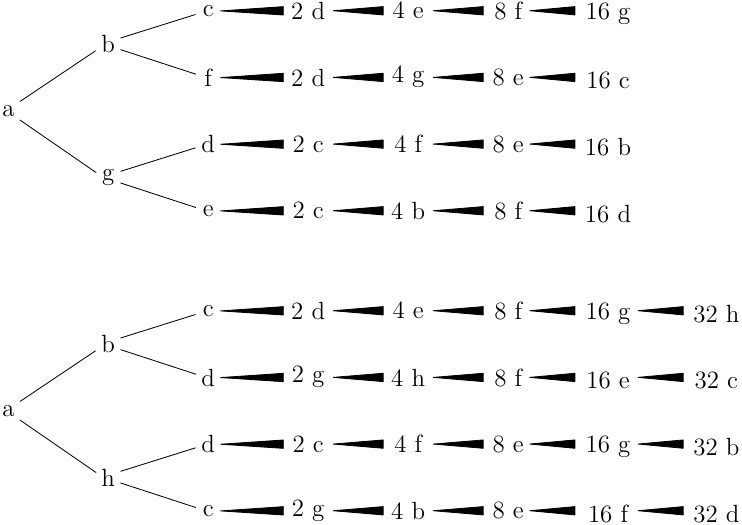}
   \caption{Top: a balanced coloring $(1,21^6)$ for a tree with height
     6. Bottom: a balanced coloring $(1,36^5, 37^2)$ for a tree with height
     7.
   }
   \label{fig:g67}
\end{figure}

Y.~Guidon~\cite{Guidon-1} gives a backtracking type heuristic method; as
his article is in french, we recall here his method.
Rather than looking for a solution by coloring the nodes one by one, his
idea consists of assembling several subtrees, each with a canonical coloring.
By grouping such subtrees, for each color $i$, we thus have a decomposition of
$a_i = q$ or $q+1$ as a sum of powers of 2.  Thus the search for a solution
is simplified by restricting the number of tests to be carried out.

For $h=3$, Guidon obtains a balanced coloring $(1,4,5,5)$ with 2 different
subtrees of height 2 with canonical coloring; see Fig.~\ref{fig:g3}.

For $h=4$, he obtains a balanced coloring $(1,7,7,8,8)$ with 4 different
subtrees of height 2 with canonical coloring; see the left tree in
Fig.~\ref{fig:g45}.

For $h=5$, he obtains a balanced coloring $(1,12,12,12,13,13)$, with 4
different subtrees of height 3 with canonical coloring, by decomposing
$q+1=13$ into $1+4+8$ and $q=12$ into $1+1+2+8$ or $2+2+4+4$; see the right
tree in Fig.~\ref{fig:g45}.

This method makes it possible to find balanced colorings for trees of
height 6 and 7; see Fig.~\ref{fig:g67}.  For $h=6$, our solution has 4
canonical subtrees of height 4, by decomposing $q=21$ into $1+4+16$,
$1+2+2+16$ or $1+4+8+8$.

Similarly for $h=7$, we have 4 canonical subtrees of height 5, by
decomposing $q+1=37$ into $1+4+32$ and $q=36$ into $1+1+2+32$, $2+2+16+16$
or $4+8+8+16$.

However for $h=8$, we did not find any solutions by this method.
Indeed, a balanced
coloring for $h=8$ is much more complicated than those for $h \le 7$.  Let
$v$ and $w$ be the two children of the root, and $i_v$ the color of $v$;
then $i_v$ must also color $q=63$ or $q-1=62$ nodes of the subtree $T_w$
attached to $w$ of height 7.  But to decompose 62 or 63 into a sum of
$2^k$, one needs at least 5 or 6 terms, therefore at least as many
canonical subtrees in $T_w$.  So for the whole tree, a balanced coloring
has at least 10 canonical subtrees.

This makes it risky to find such a solution by hand.  It is necessary to
have a systematic algorithm, to be able to program it, which is done
in the following section.

%===========================================================================

\section{Algorithm to color a tree with a given colorable partition}
%-----------------------------------------------
%
\label{sec:algorithm}

In this section, we describe an algorithm to color nodes of a perfect
binary tree, with a given colorable partition.

Let $T$ be a perfect binary tree with height $h$, colored with the minimal
number $h+1$ of colors.  The colors are labeled by integers
$i = 0, 1, \dots, h$.  Let $a_i$ be the number of nodes of $T$ colored by $i$.
We say that $A = (a_0, a_1, \ldots a_h)$ is a \emph{colorable} partition of
$T$ if $A$ respects the colorability conditions Eqs.~(\ref{sum}-\ref{ak}).
Notice that $A$ can be maximally balanced or not; the only constraint is
the colorability conditions.  Therefore, these conditions are not only
necessary but also sufficient.

Our algorithm is recursive; it is based on the recursive definition of
a perfect binary tree, $T = (r, T_b, T_c)$, where $r$ is the root,
$T_b$ and $T_c$ are perfect binary trees with height $h-1$.
In the particular case $h=0$, $T_b$ and $T_c$ are empty, then $T$ is
reduced to a single node $r$; its only colorable partition is $A = (1)$.

We saw on Section~\ref{sec:color} that we can permute the colors labels
without effects.  Also we renumber the colors so that $A$ is a non-decreasing
sequence, i.e.\ $a_0 \le a_1 \ldots \le a_h$.

Let $b_i$ and $c_i$  be the number of nodes of $T_b$ and $T_c$ respectively,
colored by $i$.
As the root $r$ of $T$ is the only node with its color, this color must
be the label 0; then $a_0=1$ and $b_0=c_0=0$.
Consequently subtrees $T_b$ and $T_c$ of height $h-1$
are colored only with colors $i$, with $1 \le i \le h$.
Counting all nodes other than the root $r$, we see that  $a_i = b_i + c_i$
for $i > 0 $.

The goal of the following algorithm is to construct $B=(b_1,\ldots b_h)$
and $C=(c_1,\ldots c_h)$, colorable partitions of $T_b$ and $T_c$
respectively; the difficulty is that $B$ and $C$ must satisfy the
colorability conditions.

Once $B$ and $C$ built, we use the algorithm again recursively
on each subtree $T_b$ and $T_c$ independently, with height $h-1$.
This will give
the colors of the two roots of $T_b$ and $T_c$, and the 4 colorable
partitions for the 4 subtrees of height $h-2$.
We continue in this way by recurrence, for $h-3$, $h-4$, \dots until we get
$2^h$ subtrees of height 0, each reduced to a single point, which are the leaves
of the initial tree $T$.  So the whole tree is colored, level by level, from
the root to the leaves.

Note that, in general, starting from an non-decreasing sequence $A$, the
sequences $B$ and $C$ are not necessarily non-decreasing.
For example, with the tree on the right of Fig~\ref{fig:h2}, from $A=(1,3,3)$
we get  $B=(1,2)$ and $C=(2,1)$.
Also the first step of the recursive procedure is a permutation of the
color labels to obtain a non-decreasing sequence
$a_0 \le a_1 \ldots \le a_h$.
Then we can assign the color 0 to the root and build $B$ and $C$.
The procedure is then recursively applied to the subtrees.
We end with the inverse permutation of the color labels, to restore the
initial labels.

Regarding the uniqueness of the solution, for the trees with
$h \le 3$, by enumerating all their colorable partitions $A$, we note that
for a fixed $A$, there is only one pair $\{B, C\}$ which checks all the
constraints.
But for larger trees, there may be several solutions. For example, for
$h=4$ with $A=(1,2,5,8,15)$, there are two solutions
$\{B, C\} = \{ (1,2,4,8), (1,3,4,7) \}$
or $\{(1,2,5,7), (1,3,3,8)\}$,
without counting the exchange between $B$ and $C$.

The algorithm as it is described below gives a single solution, chosen in a
rather arbitrary way, but this algorithm can be easily modified to
enumerate \emph{all} the colorings compatible with a given
colorable partition $A$.

As $a_0=1$ and $a_1 \ge 2$, we need to distinguish two cases, $a_1 = 2$ and
$a_1 \ge 3$ to describe the algorithm.

\subsubsection*{Algorithm for $a_1 = 2$}

Here $a_1 = b_1 + c_1 = 2$: we have to color with the color 1 two nodes
among all nodes of $T_b$ and $T_c$.
To respect the colorability conditions
the only solution is
to choose both roots of $T_b$ and $T_c$, therefore $b_1 = c_1 = 1$.

For the other colors, $i \ge 2$, the colorability conditions
give that
$a_2 \ge 4$, $a_2 + a_3 \ge 12$, i.e.\
$a_2 + \ldots + a_k \ge 2^{k+1}-4$ for $2 \le k \le h$.

The idea is to distribute the $a_i$ nodes of color $i$ in a balanced way
between the two subtrees $T_b$ and $T_c$,
by starting by $i=2$, then 3, \dots up to $h$.

If $a_i$ is an even integer, $a_i = 2 f_i$ and we therefore choose
$b_i = c_i = f_i$ which are integers.
Note that if all $a_i$ are even integers for $i \ge 1$,
the colorability conditions are verified for $B$ and $C$,
i.e.\ $b_1 + b_2 \ge 3$,
$b_1 + b_2 + b_3 \ge 7$, \dots
$b_1 + \ldots + b_k \ge 2^k-1$.  And same for $C$.
In other words, $B$ and $C$ are indeed colorable partitions for $B$ and $C$.

There is a complication
when $a_i$ is an odd integer, $a_i = 2 f_i + r_i$, with $f_i$ integer and
a remainder of the integer division $r_i=1$.
We can attribute this remainder either to $b_i$
(in this case $b_i = f_i + 1$ and $c_i = f_i$),
or to $c_i$ (in this case $b_i = f_i$ and $c_i = f_i + 1$).

One solution is to assign the nonzero remainders $r_i=1$ alternately
between the sequence of $b_i$ and that of $c_i$, starting arbitrarily by
serving the first non-zero remainder to the sequence of $b_i$.

At the end when $k = h$, as $\sum_{i=2}^h a_i$ is even,
then $\sum_{i=2}^h r_i$ is also even,
so $\sum_{i=2}^h b_i = \sum_{i=2}^h c_i $.
In other words, there have been as many remainders $r_i=1$
attributed to the sequence of $b_i$ as to the sequence of $c_i$.

Similarly, the colorability conditions for the sequence of
$a_i$, combined with the parity rules, ensure that the $b_i$ and the $c_i$
also respect these conditions.
At step $i=2$, if $a_2$ is odd, then $a_2 \ge 5$, therefore $b_2 \ge 3$ and
$c_2 \ge 2$, which satisfies the conditions.

At step $i=3$, we have $a_2+a_3 \ge 12$, and we must distinguish
according to the parity of $a_2+a_3$.
If $a_2+a_3$ is even, then $a_2$ and
$a_3$ have the same parity.
If $a_2$ and $a_3$ are both even, then $b_2 = c_2 = a_2/2$ and
$b_3 = c_3 = a_3/2$, all integers.
If $a_2$ and $a_3$ are both odd, then $b_2 = c_2 +1$ and $c_3 = b_3 +1$,
because there are two remainders $r_2=r_3=1$ to distribute alternately.
So for even $a_2+a_3$, we always have
$b_2 + b_3 = c_2 + c_3 = (a_2+a_3)/2 \ge 6$.

If $a_2+a_3$ is odd, then $a_2+a_3\ge 13$ and one of $a_2$ or $a_3$ is
even, the other odd; also a single remainder among $r_2$ or $r_3$ is not
zero and given to one of $b_i$.  Then
$b_2 + b_3 = (a_2+a_3+1)/2 \ge 7$ and
$c_2 + c_3 = (a_2+a_3-1)/2 \ge 6$.

So in all cases at step $i=3$ , the colorability conditions are verified.

These properties above hold for any $i$.
We define the partial sums
$\alpha_i = \sum_{j=1}^i a_j$,
$\beta_i  = \sum_{j=1}^i b_j$,
$\gamma_i = \sum_{j=1}^i c_j$.

With the colorability conditions, $\alpha_i \ge 2^{i+1}-2$.
If $\alpha_i$ is even,
then $ \beta_i =  \gamma_i = \alpha_i/2 \ge 2^i-1$.
If $\alpha_i$ is odd, $\alpha_i \ge 2^{i+1}-1$,
$\beta_i = (\alpha_i + 1)/2 \ge 2^i$ and
$\gamma_i = (\alpha_i - 1)/2 \ge 2^i-1$.
Consequently, $B$ and $C$ respect the colorability conditions.

These properties are valid because the algorithm distributes
the remainders $r_i=1$ alternately between the $b_i$ and the $c_i$. Thus
$\beta_i - \gamma_i = 0$ or 1, depending on
the parity of $\alpha_i$.

More precisely, the necessary and sufficient criterion is that
$|\beta_i - \gamma_i| \le 1$ at each step $i$ of the algorithm.
In other words, for a given $i$, the distribution of the nodes of the colors
$j \le i$ must be as balanced as possible between $T_b$ and $T_c$.
Thus, when $\beta_i = \gamma_i$, we have the free choice to
give the following non-zero remainder to $B$ or $C$, which will give
different solutions.

We can program the alternate distribution of the remainders $r_i=1$ with a
token that we assign alternately to $B$ and $C$.
When $B$ has the token, we code $t_B=1$ and $t_C=0$, and the next
non-zero remainder $r_i$ will be attributed to $b_i$, then the token will be
given to $C$, coded with $t_B=0$ and $t_C=1$. And so on with $C$.

To summarize, for a given colorable partition with $a_1= 2$,
$A = (1, 2, a_2 \ldots a_h)$,
the following algorithm makes $B = (b_1, \ldots b_h)$ and
$C = (c_1, \ldots c_h)$, colorable partitions of
subtrees $T_b$ and $T_c$:

\begin{verbatim}
Algorithm :

tb = 1 ; tc = 0  # initially the token is given to B
b[1] = c[1] = 1  # color 1 for the roots of $T_b$ and $T_c$
for i from 2 to h
  half = a[i] // 2  # integer division
  if a[i] even
    b[i] = c[i] = half
  else
    b[i] = half + tb
    c[i] = half + tc
    tb = 1 - tb ; tc = 1 - tc  # token exchange
\end{verbatim}

\subsubsection*{Algorithm for $a_1\ge 3$}

Contrary to the case $a_1=2$, when $a_1\ge 3$, the two roots of the
subtrees $T_b$ and $T_c$, must be of different colors.

Arbitrarily, we color the root of $T_b$ with the color 1, and the root of
$T_c$ with the color 2. As the roots are the only ones of their color in
each of their subtrees,
  \[ b_1 = 1 \ , \  c_2 = 1 \ , \]
and by complementarity,
  \[ c_1 = a_1 - 1 \ , \  b_2 = a_2 - 1 \ . \]

For a tree of height $h=2$, the only colorable partition with $a_1\ge 3$ is
$A = (1,3,3)$. The distribution above gives indeed the only solution
$B=(1,2)$ and $C=(2,1)$,
not counting the exchange of $B$ with $C$ ; see
the tree on the right of Fig~\ref{fig:h2}.

We now consider trees of height $h \ge 3$.
For color 3, we balance then $a_3 = b_3 + c_3$ nodes colored by 3, at best
between $T_b$ and $T_c$.
According to the parity of $(a_1+a_2+a_3)$,
i.e.\ $r_3 = (a_1+a_2+a_3) \ \mbox{mod} \ 2$,
we choose arbitrarily the solution such that
$(b_1+b_2+b_3) - (c_1+c_2+c_3) = r_3$.
Thereby
\begin{eqnarray*}
b_3 & = & (a_3 - a_2 + a_1 + r_3) / 2 \ , \\
c_3 & = & (a_3 + a_2 - a_1 - r_3) / 2
\end{eqnarray*}
which are integers.

Notice that at this point, we have defined $b_1$, $b_2$, $b_3$, $c_1$,
$c_2$, $c_3$.
For the distribution of the nodes of other colors, $a_i = b_i + c_i$, with
$4 \le i \le h$, we proceed in a balanced way as described in the previous
algorithm, but by starting the ``for'' loop with value $i = 4$.
The token must be correctly initialized
with $t_b = 1-r_3$ and $t_c = r_3$. In other words, if $(a_1+a_2+a_3)$ is
odd, then the remainder $r_3=1$ has been assigned to $b_3$, and the token
goes to $C$ to balance the rest of the distribution. Otherwise, when
$r_3=0$, we arbitrarily assign the token to $B$.

Note that the choice of colors 1 and 2 for the roots of subtrees $T_b$ and
$T_c$, and the distribution of color 3, are arbitrary, because there are
sometimes other solutions.
But this arbitrariness is justified because this algorithm
systematically gives two sequences $B$ and $C$ which are indeed colorable
partitions, i.e.\ they satisfy the colorability conditions.

We have found a proof, but it is complicated by the fact that the beginning
of the sequences $B$ and $C$ can be decreasing, although the sequence of
$a_i$ is non-decreasing.
Indeed, $b_2 = a_2-1$ can be greater for several indices $i$
than $b_i \sim a_i/2$, as long as $a_i < 2a_2-3$.  Same for $c_1 = a_1-1$.
% Remember that colorability conditions apply after ordering the $b_i$ and
% $c_i$ values.

Our demonstration is then a catalog according to the parities of
$a_1, a_2, a_3$. Consequently, we do not give in this article this
demonstration, more painful than technical, because we keep hoping to find
and publish a much simpler one.

With regard to the algorithmic complexity, the distribution of a colorable
partition $A$ among $B$ and $C$ is of linear order $O(h)$
for a tree of height $h$.
However, to apply the algorithm recursively to the two subtrees of height
$h-1$, the sequences $B$ and $C$ must be sorted and the colors swapped.

The global complexity will actually depend on the nature of the expected
result.  For example, if we want the complete and explicit coloring of the
tree, i.e.\ the list of colors of each of the $2^{h+1}-1$ nodes, the
complexity is $O(2^h)$, both for calculation time and memory consumption.

%===========================================================================

\section{Examples of coloring obtained with the algorithm}
\label{sec:examples}

To judge the relevance of our algorithm, we check what it gives
in the easiest case, the canonical coloring,
where all the nodes of the same height have the same color.
For a tree of height $h$, the canonical coloring partition is
$A = (1, 2, 4 \ldots 2^h)$.
We are in the case $a_1 = 2$, and all the $a_i$ are even (except of course
$a_0$ which is always equal to 1).
The algorithm gives $B = C = (1, 2, \ldots 2^{h-1})$, which corresponds well to
the canonical coloring of the two subtrees of height $h-1$, with their two
roots of same color.  By induction on $h$, the algorithm correctly
reconstructs the canonical coloring.

The goal of this work is to find coloring corresponding to balanced partitions.
For $h=2$, our algorithm gives the only balanced coloring ; see
the tree on the right of Fig~\ref{fig:h2}.
But for $h=3,4,5,6,7$, it gives different colorings than those obtained by the
method of Guidon~\cite{Guidon-1}, and shown on
Figs.~\ref{fig:g3}-\ref{fig:g67}.
The main interest of our algorithm is to give solutions for $h \ge 8$.

To illustrate the algorithm, we will detail a balanced solution for height
8 with the colorable partitions obtained at each recursive step.
We start with  the maximally balanced colorable partition $A_8$ and
the first step gives $(B_7, C_7)$ for subtrees of height 7, with
\begin{eqnarray*}
  A_8 & = & \verb+(1, 63, 63, 64, 64, 64, 64, 64, 64)+, \\
  B_7 & = & \verb+   ( 1, 62, 32, 32, 32, 32, 32, 32)+, \\
  C_7 & = & \verb+   (62,  1, 32, 32, 32, 32, 32, 32)+.
\end{eqnarray*}

For the next step, we must sort the sequences $B_7$ and $C_7$ in
non-decreasing order, swapping the color labels.
Here $B_7$ and $C_7$ gives the same sorted sequence $A_7$, but with
different permutations. Then
\begin{eqnarray*}
  A_7 & = & \verb+(1, 32, 32, 32, 32, 32, 32, 62)+, \\
  B_6 & = & \verb+   ( 1, 31, 16, 16, 16, 16, 31)+, \\
  C_6 & = & \verb+   (31,  1, 16, 16, 16, 16, 31)+.
\end{eqnarray*}
The following steps give
\begin{eqnarray*}
  A_6 & = & \verb+(1, 16, 16, 16, 16, 31, 31)+, \\
  B_5 & = & \verb+   ( 1, 15,  8,  8, 16, 15)+, \\
  C_5 & = & \verb+   (15,  1,  8,  8, 15, 16)+, \\
  \\
  A_5 & = & \verb+(1,  8,  8, 15, 15, 16)+, \\
  B_4 & = & \verb+    (1,  7,  8,  7,  8)+, \\
  C_4 & = & \verb+    (7,  1,  7,  8,  8)+, \\
  \\
  A_4 & = & \verb+(1, 7, 7, 8, 8)+, \\
  B_2 & = & \verb+   (1, 6, 4, 4)+, \\
  C_3 & = & \verb+   (6, 1, 4, 4)+, \\
  \\
  A_3 & = & \verb+(1, 4, 4, 6)+, \\
  B_2 & = & \verb+   (1, 3, 3)+, \\
  C_2 & = & \verb+   (3, 1, 3)+, \\
  \\
  A_2 & = & \verb+(1, 3, 3)+, \\
  B_1 & = & \verb+   (1, 2)+, \\
  C_1 & = & \verb+   (2, 1)+.
\end{eqnarray*}

In the example above, we see that at each step, $B_i$ and $C_i$ give, once
sorted, the same sequence $A_i$.  But that is not true in general.  For
example, for the maximally balanced colorable partition of $h=9$, we obtain
according to the steps, between 1 and 4 distinct sorted sequences $A_i$.

Notice that sequences $A_4$ and $A_2$ are maximally balanced colorable
partitions, but the others are not.

\begin{figure}[p]
   \centering
   \includegraphics[width=\textwidth]{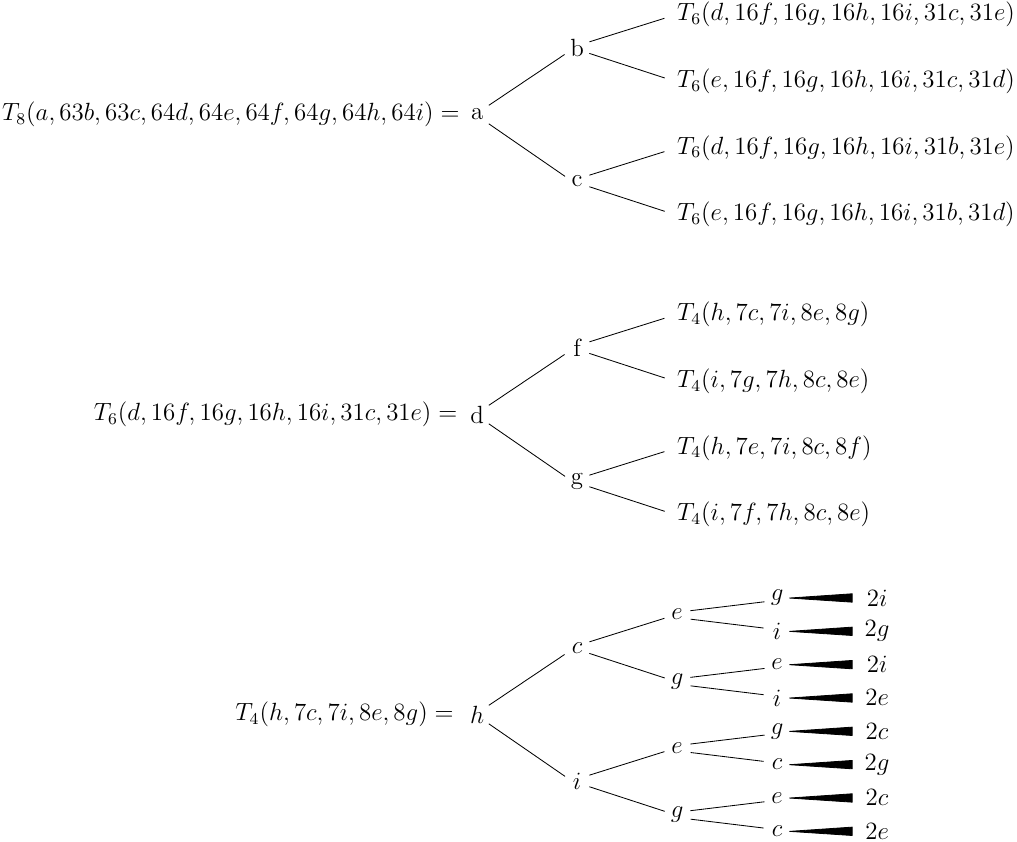}
   \caption{A maximally balanced coloring for a tree of height 8.
      See text for notations.
   }
   \label{fig:g8}
\end{figure}

The colored tree of height 8, corresponding to the solution above, is shown
in Fig.~\ref{fig:g8}.
The 9 colors are labeled by letters $a,b\dots, i$ and a notation like
$16g$ means that 16 nodes are colored with $g$.
As it is difficult to drawn a tree with 511 nodes
on one page, and to verify the good sharing of colors,
we cut it into three slices.
First the full tree $T_8$ is decomposed into 4 subtrees $T_6$ of height 6.
As all $T_6$ have the same colorable partition $A_6$, only one $T_6$ is
drawn and decomposed into 4 subtrees $T_4$ of height 4.  The other $T_6$
are obtained by changing the labels of colors.
Finally as all $T_4$ share the same colorable partition $A_4$, only one
$T_4$ is explicitly drawn.
In each slice, the reader can check that the sums of the colors on the
right give the correct numbers on the left.

With a computer, one can of course obtain solutions for larger trees, $h > 8$.
Because of the exponential complexity $O(2^h)$ of a binary tree, solutions
are difficult to draw on a sheet of paper.  Similarly, the saturation limit
of a computer's memory is reached quickly.

%===========================================================================

\section{Conclusion}
%------------------

To answer the problem exposed by Guidon~\cite{Guidon-1}, a balanced
coloring of a perfect binary tree, we presented a systematic algorithm
which gives a solution (and possibly all solutions) for a tree of
arbitrarily large height; the only limits are the power of the computer and
the size of its memory.

This algorithm is designed to balance, from each node, the colors between
its two subtrees.  For a given color, the set of nodes of this color is
generally rather spatially fragmented, both in height and in the width of
the tree.  See for example, the solution for height 8 solution exposed in
Fig.~\ref{fig:g8}.

This contrasts with Guidon's method which, on the contrary, tries to group
nodes of the same color on the same height to facilitate the search for a
solution.
Guidon was initially motivated by the wiring of a control system for
a network of tree switches: he wanted to minimize the number of commands
(here, the number of colors) and the maximum power of a command (here, the
number of nodes of a given color).

We see that the fragmentation of colors, inherent to our algorithm, can
make the wiring of the control system too complex.
The solution given by our algorithm is somehow \emph{too} balanced.  Also
we could introduce into the minimization criteria, in addition to the color
balance, a variable which favors the spatial grouping of nodes of the same
color.  However, our algorithm will be difficult to generalize to this new
problem, because we lose the decoupling between subtrees, which is the
basis of our induction algorithm.

\subsection*{Acknowledgments}
%-----------------------------

It is a pleasure to thank Y.~Guidon for a stimulating discussion and for
sharing Ref.~\cite{Guidon-2} before publication.  This research did not
receive any specific grant from funding agencies in the public, commercial,
or not-for-profit sectors.

%===========================================================================

\section*{Appendix: number of colorings of a tree}
%---------------------------------------------

In this appendix, we give an exact formula for the number of colorings of a
tree, and its asymptotic expansion for large trees.

Let $c_n$ be the number of colorings with $n$ colors of a perfect binary tree
of height $n-1$, with the coloring rule defined Section~\ref{sec:rule}.

To define $c_n$, labels of color do not matter as for graph theory; we can
change or permute the labels without effects.
But it is easier to calculate $d_n$, the number of colorings where we
discern the labels of the $n$ colors.
Thus each
coloring counted by $c_n$ corresponds to the $n!$ colorings counted by
$d_n$, associated with the $n!$ permutations of the $n$ color labels. Thereby
\begin{equation}
  d_n= n! \: c_n .  \label{dncn}
\end{equation}

To count $d_n$, we start by coloring the root of $T$: we can freely choose its
color, so there are $n$ choices.
Because of the coloring rule, the color of the root can no longer be used
to color nodes of the two subtrees $T_L$ and $T_R$ hooked to the root;
we must therefore color them with the remaining $n-1$ colors.
To respect the coloring rule globally on the tree $T$, it suffices to
respect the coloring rule independently on $T_L$ and $T_R$ with the $n-1$
colors, i.e.\ without the color of the root of $T$.
As $T_L$ and $T_R$ are trees of height $n-2$ colored with $n-1$ colors,
there are $d_{n-1}$ possible colorings for $T_L$, and also $d_{n-1}$ for
$T_R$.  Thus

\begin{eqnarray}
  d_n & = & n \: d_{n-1}^2,    \label{recur_d} \\
  c_n & = & (n-1)! \: c_{n-1}^2.
\end{eqnarray}
The recurrence~(\ref{recur_d}) is sometimes called
``Somos's quadratic recurrence''~\cite[page 446]{Finch}

Starting with $d_1=1$, the sequence of $d_i$ begins with
$(1, 2, 12 , 576, 1658880 \ldots)$ which is the sequence A052129
in OEIS~\cite{OEIS}.
Similarly, with  $c_1$ and $c_2=1$, the sequence of $c_i$ begins with
$(1, 1, 2, 24, 13824, 22932357120,\ \ldots)$.
Using Eqs.~(\ref{recur_d}) and (\ref{dncn}),
\begin{eqnarray}
  d_n & = & n (n-1)^2 (n-2)^4 \dots 2^{2^{n-2}}
        = \prod_{k=1}^n k^{2^{n-k}}, \\
  c_n & = & (n-1) (n-2)^3 \dots 2^{2^{n-2}-1}
        = \prod_{k=1}^{n-1} k^{2^{n-k}-1}.    \label{cn}
\end{eqnarray}

To obtain the asymptotic behavior at large $n$, we need the
\emph{quadratic recurrence constant}~\cite{Somos}\cite[page 446]{Finch},
\begin{equation}
    U = \sqrt{1 \sqrt{2 \sqrt{3 \sqrt{4 \sqrt{\ldots}}}}}
      = \prod_{k \ge 1} k^{1/2^k}
      = 1.6616879496\ldots  \label{u}
      % 1.6616879496335941212958189227499507499644186
\end{equation}
whose digits form the sequence A112302 in OEIS~\cite{OEIS}.

Considering $U^{2^n} = \prod_{k\ge 1} k^{2^{n-k}}$,
the product is cut into two parts, $k\le n$ (which gives $d_n$) and $k>n$; then
$U^{2^n} = d_n \, r_n$ where
\begin{equation}
  r_n = \prod_{k \ge 1} (n+k)^{1/2^k} = (n+1)^{1/2} (n+2)^{1/4}(n+3)^{1/8}\ldots
\end{equation}
By dividing each factor of $r_n$ by $n$,
%knowing that $\sum_{k\ge 1} 1/2^k = 1$,
\begin{equation}
  r_n = n (1+\frac{1}{n})^{1/2} (1+\frac{2}{n})^{1/4} (1+\frac{3}{n})^{1/8}\ldots
\end{equation}

Then $d_n = n! c_n = U^{2^n} f(1/n)/n$ where
\begin{equation}
   f(x) = (1+x)^{-1/2} (1+2x)^{-1/4}(1+3x)^{-1/8} \ldots
        = \prod_{k \ge 1} (1+kx)^{-1/2^k} \ .
\end{equation}
As $f(x)$ is continuous at $x=0$ and $f(0)=1$, the asymptotic behavior for
large $n$ is
\begin{equation}
    d_n \sim \frac{U^{2^n}}{n} \quad ; \quad c_n \sim \frac{U^{2^n}}{n \: n!}.
\end{equation}

The asymptotic expansion of $f(1/n)$ can be obtained by considering
\begin{equation}
    g(x) = \log f(x) = - \sum_{k \ge 1} \log(1+kx)/2^k.
\end{equation}
By expanding the log, we get
\begin{equation}
  g(x) = \sum_{i \ge 1} (-1)^i S_i \, x^i/i
\end{equation}
with
$S_i =  \sum_{k \ge 1} k^i/2^k $.
These numbers are integers; starting with $S_1=2$, the sequence of $S_i$
begins with $(2, 6, 26, 150, 1082, 9366, ...)$, sequence A000629 or A076726
in OEIS~\cite{OEIS}.

So the expansion of $f(x) = \exp(g(x))$ begins with
\begin{equation}
    f(x) = 1 - 2x + 5x^2 - 16x^3 + 66 x^4 - 348 x^5 + ...
\end{equation}
where the coefficients form the sequence A084785 of OEIS~\cite{OEIS}.

%==============================================================================

\end{document}